\documentclass[iop,apj]{emulateapj}
\usepackage{apjfonts}
\usepackage{graphicx}
\usepackage{amsmath,amstext}
\usepackage{tikz}
\usepackage{enumitem}
\usepackage[flushleft]{threeparttable}
\usepackage{color}
\usepackage{hyperref}
\usepackage{natbib}
\usepackage[titletoc,title]{appendix}
\definecolor{darkblue}{rgb}{0.0,0.0,0.4}
\hypersetup{colorlinks,
            linkcolor=darkblue,urlcolor=darkblue,
            citecolor=darkblue}

\newcommand{\angstrom}{\,\textup{\AA}}
\newcommand{\lya}{Ly$\,\alpha$ }

\newcommand{\quasar}{Q\,0420$-$388 }

\shorttitle{Quasar lifetimes constraints using fluorescence}
\shortauthors{Borisova et al.}

\begin{document}

\title{Constraining the lifetime and opening angle of quasars using fluorescent Lyman {\Large$\alpha$} emission: the case of Q0420-388}
\author{Elena Borisova\altaffilmark{1,*}, Simon J. Lilly\altaffilmark{1}, Sebastiano Cantalupo\altaffilmark{1}, J. Xavier Prochaska\altaffilmark{2}, Olivera Rakic\altaffilmark{3}, Gabor Worseck\altaffilmark{3}}

\affiliation{$^1$Institute for Astronomy, ETH Zurich, Zurich, CH-8093, Switzerland}
\affiliation{$^2$UCO/Lick Observatory, UC Santa Cruz, Santa Cruz, CA-95064, USA}
\affiliation{$^3$Max-Planck-Institut f\"{u}r Astronomie, Heidelberg, D-69117, Germany}

\email[*]{borisova@phys.ethz.ch}

\begin{abstract}

A toy model is developed to understand how the spatial distribution of fluorescent emitters in the vicinity of bright quasars could be affected by the geometry of the quasar bi-conical radiation field and by its lifetime.  The model is then applied to the distribution of high equivalent width Lyman~$\alpha$ emitters \citep[with rest-frame equivalent widths above 100~$\angstrom$, threshold used in e.g.][]{Trainor} identified in a deep narrow-band 36x36~arcmin$^2$ image centered on the luminous quasar \quasar. These emitters are found to the edge of the field and show some evidence of an azimuthal asymmetry on the sky of the type expected if the quasar is radiating in a bipolar cone.  If these sources are being fluorescently illuminated by the quasar, the two most distant objects require a lifetime of at least 15~Myr for an opening angle of 60~degrees or more, increasing to more than 40~Myr if the opening angle is reduced to a minimum 30~degrees. However, few of the other expected signatures of boosted fluorescence are seen at the current survey limits, e.g. a fall off in Lyman~$\alpha$ brightness, or equivalent width, with distance.  Furthermore, to have most of the Lyman~$\alpha$ emission of the two distant sources to be fluorescently boosted would require the quasar to have been significantly brighter in the past.  This suggests that these particular sources may not be fluorescent, invalidating the above lifetime constraints. This would cast doubt on the use of this relatively low equivalent width threshold and thus on the lifetime analysis in \citet{Trainor}.

\end{abstract}

\maketitle
\newpage

\section{Introduction}

Neutral gas that is illuminated by ultraviolet radiation above the Lyman limit will fluorescently emit H\,I \lya photons \citep{Hogan,Gould}. Although the surface brightness levels expected from the ambient ultraviolet radiation field are extremely low, illumination by a nearby bright source can boost this fluorescent emission to detectable levels \citep{Haiman-Rees-2001, C2005, 2010Kolmeier}. This opens up the possibility of mapping the filaments of the cosmic web, detecting early gas rich phases of galaxy formation, and also, as explored in this paper, establishing constraints on the emission geometry and luminous history of the quasars.

Over the last few years, there has been growing observational evidence for fluorescent emission in the neighbourhood of bright quasars~\citep{Adelberger, Kashikawa}. \citet{C2007q} presented a number of plausible fluorescent candidates that were found in a multi-slit survey around quasar \quasar at z=3.110. Signatures of such emission include high \lya equivalent width, double peaked line profiles and a surface brightness that should be simply related to the incident radiation. However none of these are unambiguous indicators of fluorescence and all can be modified by non-ideal gas geometries and/or velocity structure \citep{C2005,C2007q}.

More recently \citet{C2012} obtained a very deep narrow-band image of the high luminosity quasar HE0109-3518 at $z = 2.4$. This contained at least 18 objects with rest-fame \lya equivalent widths well above the limit of 240$\angstrom$ that can be expected from photo-ionization by stellar populations \citep{ew240_Charlot,Schaerer2003}. In fact, a stack of 12 of these objects had an impressive equivalent width exceeding EW$_0\ge 800\angstrom$. These objects were interpreted as gas rich ``dark galaxies'' which were illuminated by the quasar radiation.

As pointed out by \citet{C2005,C2008}, if fluorescent emission can be convincingly detected around quasars, then the spatial distribution of such emission can be used to set constraints on both the quasar lifetime (how long it has been shining at similar luminosity) and on whether the quasar emission is isotropic and, if so, on the illumination geometry. Fluorescent \lya emission is the (three-dimensional) emission equivalent to the (one-dimensional) ``proximity effect'' seen in the decrease in the number of \lya absorption systems per unit redshift in the neighbourhood of a bright quasar \citep{Proximity1988,Proximity2000}.

In a recent large narrow-band survey and follow-up spectroscopy of eight fields centered on hyper-luminous QSOs at $2.5<z<2.9$, \citet{Trainor} have found 116 objects with rest-frame equivalent widths above $100\angstrom$ with $32$ above $240\angstrom$ threshold. Taking this lower $100\angstrom$ limit as a signature of fluorescent emission, \citet{Trainor} put constraints on the opening angles and lifetimes of the observed quasars. From the distribution of those objects in redshift and in the plane of the sky they concluded that most of quasars in their sample have been shining with the comparable luminosity for a time $1\,$Myr$~\lesssim~t_Q ~\lesssim~20\,$Myr within an opening angle $\sim 30\degr$ or larger. Because their fields were quite small, most information is contained in the distribution of objects along the line of sight, i.e. specifically an asymmetry in front of and behind the quasar position due to the effects of light travel time  for the background objects. However, simple light travel time effects may be significantly modified and made more complicated if quasars shine intermittently or if they have asymmetrical (e.g. unipolar) emission.

In this paper we develop a multi-parameteric toy model to gain an understanding of geometrical and light travel effects that could be expected on the distribution of fluorescent objects when a large three-dimensional volume is illuminated by bi-conical quasar radiation~\citep{Antonucci,1995U}.  We focus on the projected distribution of the emitting sources in the plane of sky as it is relevant for sources detected in narrow-band images. 

We then apply these models to the observed distribution of high equivalent width \lya emitters that have been detected in a deep narrow-band survey of a relatively large field (at least in comparison with previous studies) around the bright quasar \quasar at $z = 3.110$.  A total of 264 \lya emitting (LAEs) candidates were detected, amongst which 16 have rest frame equivalent widths above $100\angstrom$, the limit which was used to isolate fluorescence by \citet{Trainor}. These sources extend across the full $36\times 36$ arcmin$^2$ field and show some evidence for a non-uniform distribution on the sky.  This is seen relative to that expected for an azimuthally uniform distribution and also relative to the observed distribution of lower equivalent width \lya emission line sources. The azimuthal asymmetry in the distribution matches quite well that expected for a bi-conical illumination from the quasar.   If we assume that these high equivalent width sources are being fluorescently illuminated, then we can already set significant constraints on the quasar lifetime and opening angle. On the other hand, we do not see some of the expected signatures at the current survey limits, including a radial fall off in \lya luminosity or EW.  Furthermore, the most distant sources would require the quasar to have been significantly brighter in the past if most of the \lya is fluorescently reradiated quasar emission.  We note that this 100$\angstrom$ limit as an observational threshold for fluorescence which is substantially lower than the $240\angstrom$ EW cut, adopted in~\citet{C2007q}. This threshold is discussed in more details further below in Section~\ref{subsec: highew}.

\begin{figure*}[t]
\includegraphics[width=\textwidth]{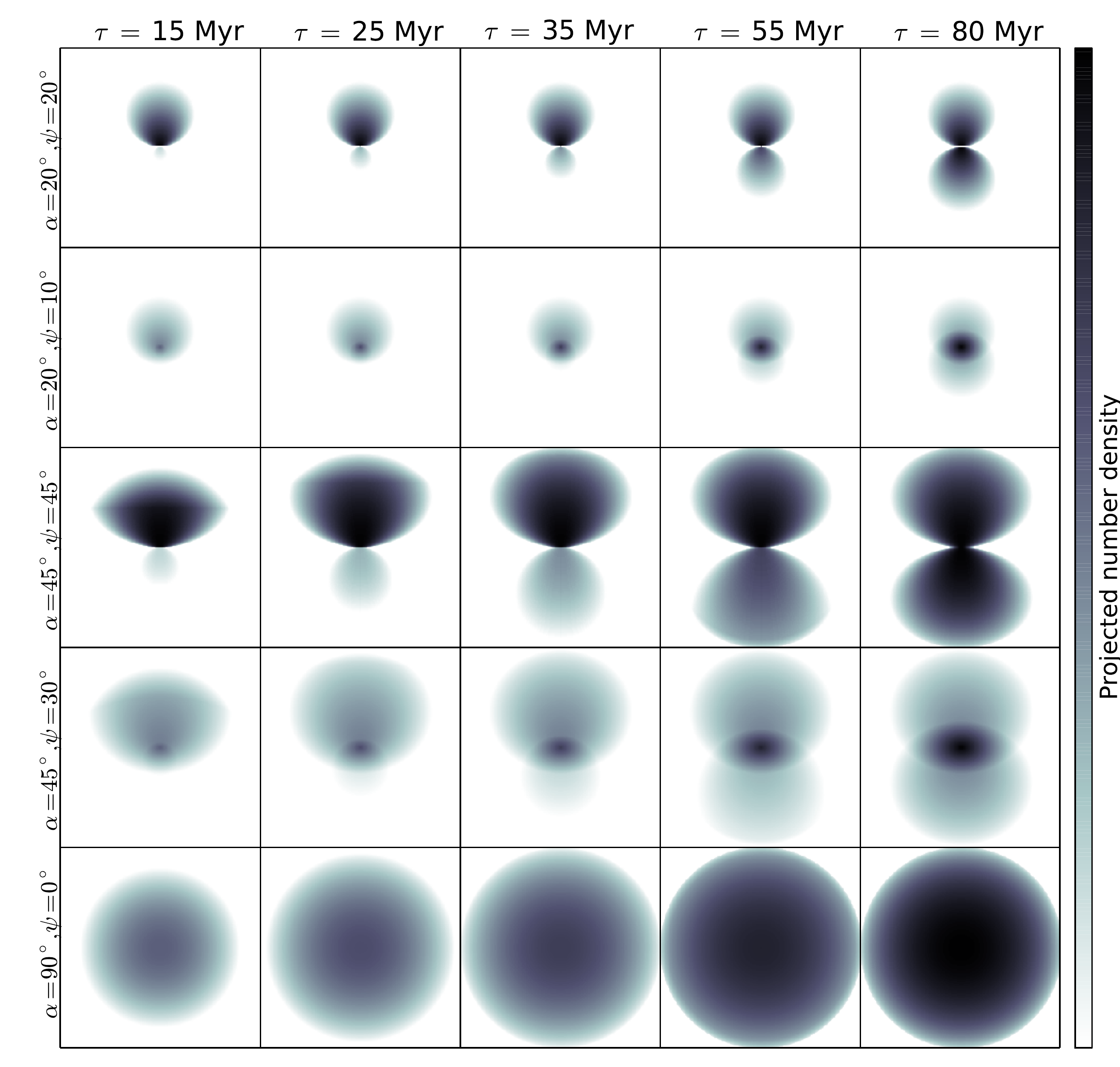}
 \caption{The projected surface density of objects uniformly distributed in space and illuminated by the quasar for different sets of opening and viewing angles (labeled respectively $\alpha$ and $\psi$) and different quasar lifetimes $\tau$. The side of each square corresponds to $r=10\,$pMpc. Each horizontal raw shows cases of different pairs of $(\alpha,\psi)$ while columns are snapshots at different ages of the quasar. The color scale in surface density is renormalized for each row. See text for description of the different effects that are visible.}
 \label{figgrid}
\end{figure*}

The structure of the paper is as follows. In Section~\ref{sec: genarg} we develop an intuitive feeling for what the distribution of the fluorescently boosted emitters around the quasar would look like given the orientation of bi-conical radiation and the lifetime of the quasar. The model is parameterized by an opening angle and parameters describing the quasar lifetime and the maximum distance to which quasar-related fluorescence can be detected plus observational parameters such as the width of the narrow-band filter used in \lya surveys. In Section~3 we apply our model to the real data. In Sections~\ref{subsec: observations}-\ref{subsec:catalog} we describe how the data was obtained and reduced and what algorithm we used to select \lya emitters. In Sections~\ref{subsec: highew}-\ref{subsec: genconstraints} we discuss the significance of the apparent asymmetry in the distribution of high-EW objects, selected above $100\angstrom$ threshold. Finally, in Section~\ref{subsec: maxlikelihood} we do a maximum likelihood fit of these models to the observed distribution of high-EW objects to derive constraints on the quasar emission parameters, subjected to the assumption that they are indeed being fluorescently illuminated.

Throughout the paper we assume a concordance cosmology with H$_0=70\,$km$\,$s$\,^{-1}\,$Mpc$^{-1}$ and $(\Omega_m,\Omega_\Lambda)=(0.3,0.7)$.

\section{Fluorescence as a diagnostic of lifetime and opening angle}
\label{sec: genarg}

\subsection{General arguments}

\begin{figure*}[ht]
\includegraphics[width=\textwidth]{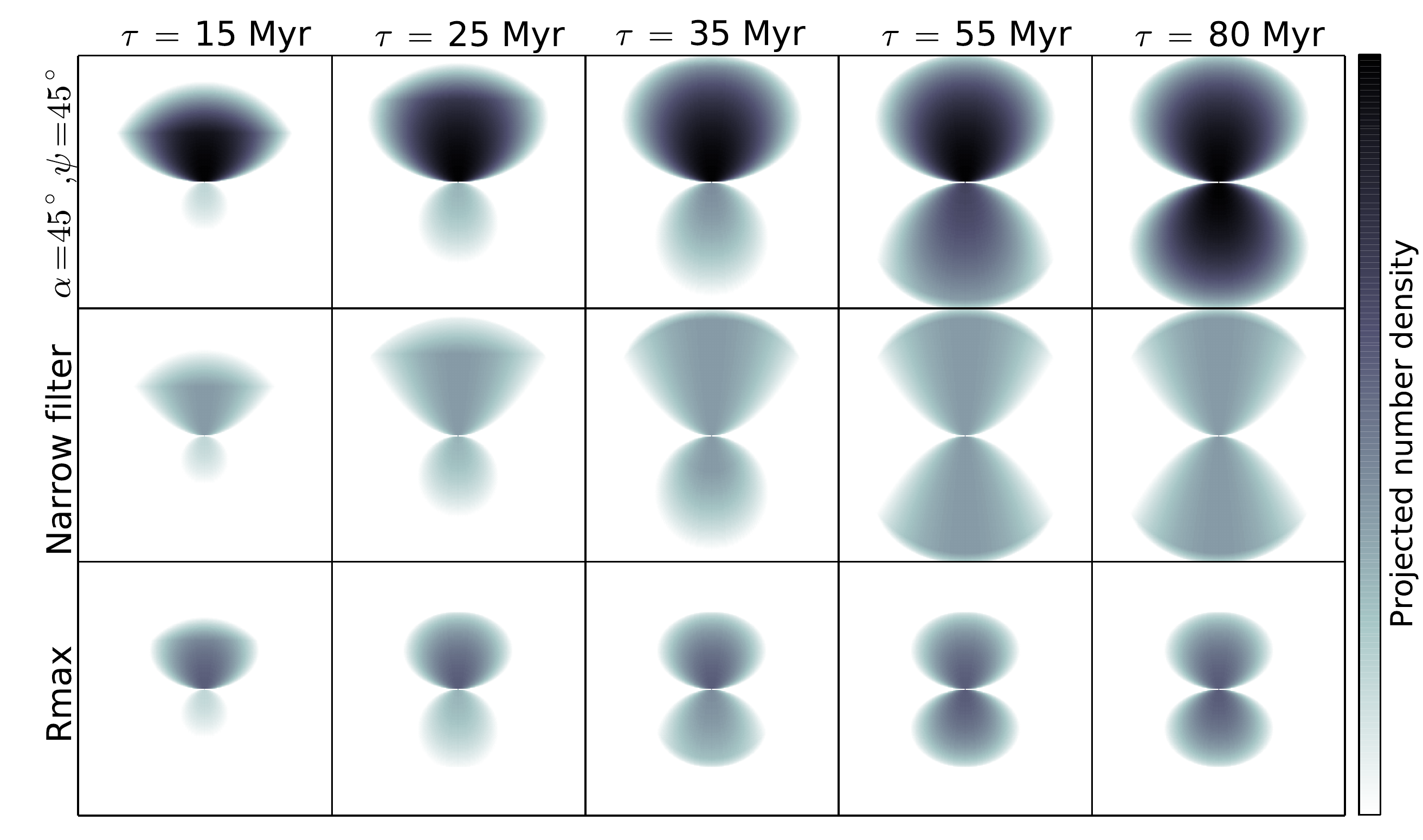}
 \caption{As for Fig 1, but illustrating the effects introduced by a narrow-band filter and by a limited zone of influence of the quasar boosting. The normalization is now common for all the panels. The first row is the same as the middle raw in Fig.~\ref{figgrid}, the second row introduces the presence of a finite narrow-band filter, and the third row demonstrates how the picture changes if the quasar boosting is limited to a fixed three-dimensional distance from the quasar.}
 \label{figgrid-filter}
\end{figure*}

The spatial distribution of those objects for which it can be argued that the \lya emission has been boosted by quasar radiation can give constraints on the lifetime and opening angle of the quasar emission. There will be a time delay between the moment when an observer on the Earth will see the quasar turn on and when he will see the response of neutral gas in the neighbourhood of the quasar. The surfaces of constant time delay will be paraboloids of rotation aligned with the line of sight and with the quasar at the focus:

$$
x^2 + y^2 - 2 c\tau(z+\frac{c\tau}{2}) = 0, 
$$

\noindent
where $\tau$ is the quasar lifetime and $c$ - speed of light. The z-axis coincides with the line of sight and $(x,y)$ define the plane of the sky.

In the simplest case of an instantaneous turn-on of the quasar, fluorescent emission will not be observed outside of the region defined by the paraboloid corresponding to a time delay equal to the elapsed time since turn-on. The fact that we see the quasar still shining then implies a minimum quasar lifetime. Since the boosting effect of the quasar radiation falls off with distance, fluorescent emission will also not be seen outside of a spherical region centered on the quasar whose radius corresponds to the maximum distance from the quasar for a significant boost. The boosting factor $b$, which describes how the fluorescence radiation is amplified by quasar radiation field in respect to the ambient level, falls with the distance $r$ as $b\sim r^{-2}$ but, as was shown in \citet{C2007q}, in the case of very bright quasars can still be significantly high at large distances ($b$ may exceed a factor of two at distances of tens of comoving Mpc (cMpc) for very bright quasars). Finally, the volume over which we can see fluorescent emission will also be limited to the slab in space corresponding to the range of \lya redshifts admitted by the narrow-band 
filter.

In the analysis of this paper, we will consider the simple case in which the quasar turns on at some point in the past and has maintained a more or less constant luminosity (and orientation) since then. We have assumed that \lya emission occurs instantaneously, i.e.\ we neglect the recombination time scale. The recombination time depends on the density of the medium. For the typical densities of fluorescently illuminated gas at the cloud luminosities probed here, i.e. $n > 0.1\,$cm$^{-3}$ \citep[see e.g.][]{C2005,C2012}, this contribution is probably negligible but for low density clouds ($n_{HII}\sim0.01$\,cm$^{-3}$) this additional delay could be as high as $\sim10\,$Myr. If the recombination time is significant, it could be added to the quasar lifetime that is derived from geometric arguments. 

We will also assume that the quasar emits radiation only within two solid cones that share a common axis.  This bipolar emission will therefore be characterized by a \textit{semi-opening} angle $\alpha$.  Since we see the quasar ourselves, we will only consider cases in which the bipolar axis is oriented such that our line of sight to the quasar lies within one of these two cones.  Spherically symmetric quasar emission is represented by $\alpha$ approaching 90$^\circ$.

\subsection{Toy model: density of fluorescent objects}
\label{subsec:toy}

In order to visualize the spatial distribution of illuminated objects, a simple geometrical toy model was developed. In a given volume the model computes which regions are illuminated by the quasar bi-cone radiation and projects these onto the plane of the sky. If the space density of potentially fluorescent sources is uniform around the quasar, then this projected volume will describe the expected surface density distribution of fluorescent sources. The illuminated volume consists of the co-axial double cone, bounded by the paraboloid surface set by the lifetime $\tau$ of the quasar (i.e. the time since the onset of illumination) and also a spherical shell reflecting the maximum distance $r_{max}$ to which significant boost fluorescence can be produced.  The orientation of the cone in space is set by two angles: $\phi$, the projected position angle of the axis of the cone on the sky, and $\psi$, the angle between the line of sight and the axis of the cone.  The fact that we see the quasar, i.e. we are looking down one of the illumination cones, implies $|\psi| \le \alpha$.

The results for different representative input parameters are shown in Fig.~\ref{figgrid}. The $\phi$ angle was chosen in these plots to be  $0^\circ$ for simplicity, so that the cones are oriented along the vertical axis, and $r_{max} = 10\,$physical Mpc (pMpc). The five columns correspond to increasing quasar lifetimes, left to right. The different rows correspond to different pairs of the $(\alpha,\psi)$ parameter, as indicated on the left side, i.e. on the half-opening angle of the bipolar cone and on where within the cone the line of sight passes. The normalization is common within each row of snapshots but varies from row to row.  Several generic features of the projected distributions are readily apparent on these plots.

First, the true bi-conical asymmetry is most clearly seen only for a small range of viewing angles with $\psi \sim \alpha$ (see e.g. rows 1 and 3 in Fig.~\ref{figgrid}), i.e. the relatively unlikely configuration where our line of sight penetrates down to the quasar close to the surface of the cone.  In this case, there is a characteristic ``figure of eight'' pattern with a ``zone of avoidance'' oriented perpendicular to the projected axis of the cone.  As soon as the axis of the cone is rotated closer to the line of sight, i.e. $\psi < \alpha$ this zone of avoidance is filled in and the surface density distribution peaks at the quasar due to the overlapping of the two cones, which is shown in row 2.   Second, light travel time effects produce a strong asymmetry between the two lobes.  This persists until the time when the quasar has shone for long enough ($\tau\sim  50\,$Myr) that the illuminated volume (as seen by us) is defined by the spherically symmetric $r_{max}=10\,$pMpc rather than by the paraboloid time-delay surface.  Finally, the case of isotropic quasar emission, shown at the bottom row, always produces an azimuthally symmetric surface density distribution.

A common technique for \lya searches is based on narrow-band observations centered on blind fields or on a bright quasar. To see the effect that the narrow-band filter places in the redshift space, we bound the illuminated volume by two parallel surfaces (perpendicular to the line of sight) that represent the redshift limits associated with the passband of the $50\angstrom$ narrow-band filter. This is shown in Fig.~\ref{figgrid-filter}, which has the same normalization as Fig.~\ref{figgrid}. The top row of Fig.~\ref{figgrid-filter} reproduced the middle row of  Fig.~\ref{figgrid}, i.e. an opening angle $\alpha = 45^\circ$ seen when the observer is looking down the surface of the cone $\psi = 45^\circ$. The second row in Fig.~\ref{figgrid-filter} shows the effect of a filter with a typical width of narrow-band filter used in \lya surveys.  It can be seen from the number density of projected sources that the maximum surface density is reduced, the zone of avoidance has become slightly larger and the equalization of the two lobes happens earlier (i.e. after a shorter quasar lifetime). 

In the third row of Fig.~\ref{figgrid-filter} we reduce the radius $r_{max}$ . As would be expected, this also decreases the lifetime at which the two lobes become symmetrical while also severely limiting the projected size of the distribution.

These two figures give a sense of the features of the projected spatial distribution of fluorescently boosted objects that might be expected to be seen around bright quasars.   It can be seen that there is information in the maximum projected distance from the quasar, in the azimuthal distribution of sources and in particular the possible existence of a narrow ``zone of avoidance'', and in the asymmetry between the two lobes.

\section{Model application: the case of \quasar}

Deep narrow-band imaging can detect large numbers of high equivalent width emission line sources. These are dominated by Lyman $\alpha$ emitters at high redshifts.  We have obtained a deep wide field narrow-band image of the bright quasar \quasar at the redshift of $z=3.110$. It represents a unique data set to search for the signatures of quasar radiation geometry and lifetime. In this section we apply the results of Section~\ref{sec: genarg} to a particular set of data. We first provide the details on how the data was obtained and reduced and then how the candidates of fluorescent \lya emission boosted by the quasar were selected.

We compare the observed distribution of high equivalent emitters with the prediction of our toy model.  We stress at the outset that the identification of the high equivalent width emitters in this field as being fluorescently boosted by the quasar emission is not at this point secure.  Subsequent spectroscopy would be required to verify the nature of \lya emitters and to add a third dimension to the model comparison. 

\subsection{CTIO imaging observations}
\label{subsec: observations}

\begin{figure}[t!]
\includegraphics[width=0.5\textwidth]{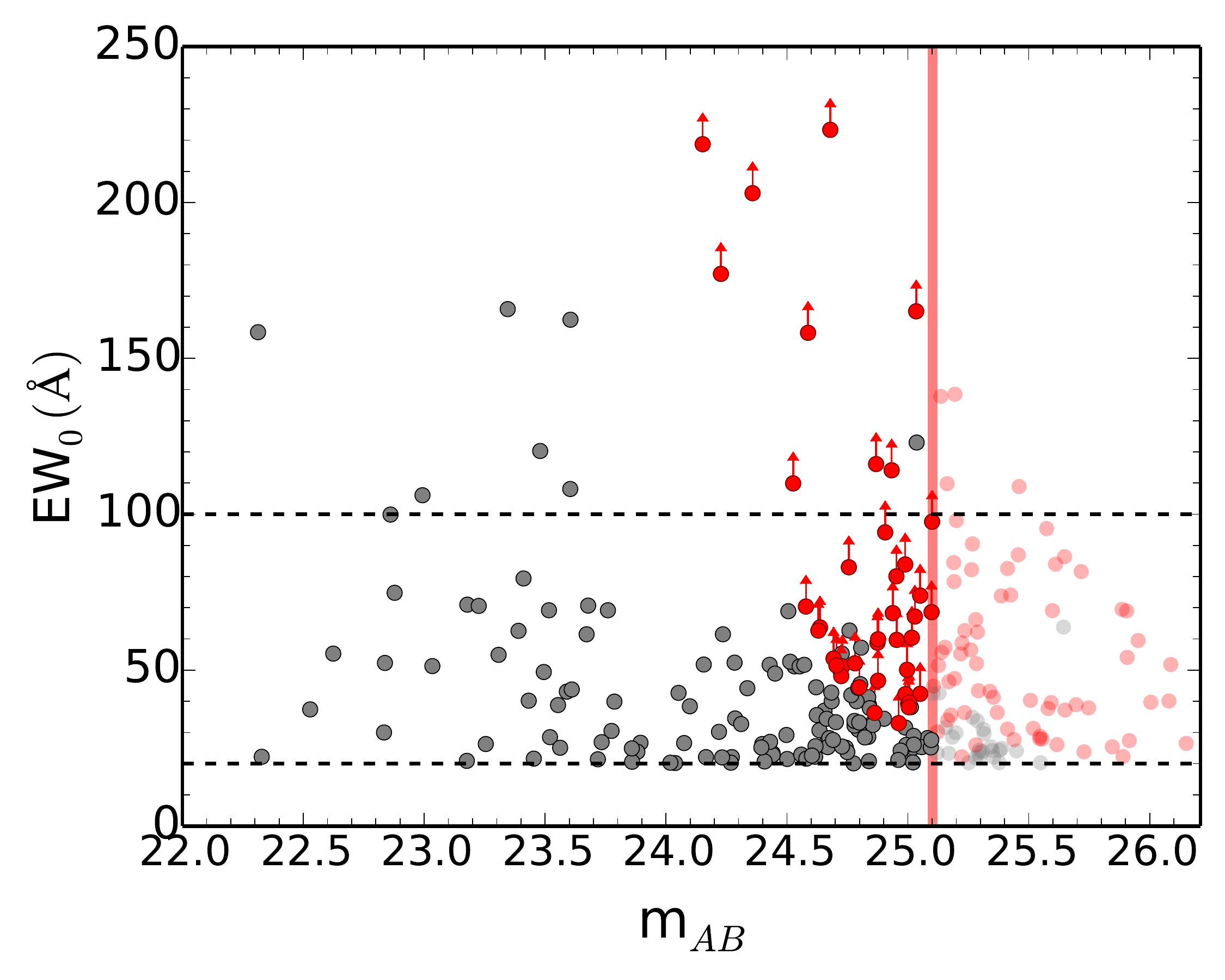}\caption{Narrow-band magnitudes of LAE candidates versus the rest-frame equivalent width derived from the photometry (assuming the line is \lya). The lower horizontal line at $20\angstrom$ shows the global cut for \lya emitters, the upper horizontal line at $100\angstrom$ is the assumed upper limit for \lya line emission from star forming objects. The vertical line represents the mean completeness limit for source detection across the CCDs in the mosaic estimated by adding point sources to the image and running the detection algorithm. The red points show objects with continuum level below $3\,\sigma$ and which therefore have only lower limits on their equivalent width.}
\label{magew}
\end{figure}

Narrow and broadband images of a large field centered on the $z=3.110$ luminous quasar \quasar were obtained over six consecutive nights from 22 to 27 of November 2011 using the wide field MOSAIC II camera mounted on the CTIO Blanco 4m telescope. The camera consists of eight 2048x4096 CCDs which cover the area of 36x36 arcmin$^2$. The seeing conditions varied from 0.90 to 1.30 arcsec. Broadband imaging was obtained in the $u,g$ and $r$ filters. The narrow-band images were taken through a $50\angstrom$ wide filter (FWHM) with central wavelength at $\lambda=4990\angstrom$ which includes \lya over about $\sim$10~pMpc in the redshift dimension at the redshift of $z\approx3.1$. A dithering pattern was used in order to cover the gaps ($\approx 18"$) between the CCDs in the final combined image. The characteristics of the filters and total exposure times for each filter are listed in Table~\ref{filttable}.

The redshift of \quasar is somewhat uncertain because of the broad lines in the spectrum and also due to the general fact that high-ionization emission lines are frequently blueshifted with respect to the systemic redshift~\citep[e.g.][]{ZblueShift,Shen2016}. Different values of the redshift are found in the literature. \citet{z3.110_1994} derived the value $z=3.110$, which was used by \citet{C2007q} in a ``multislit plus filter'' survey for fluorescent emission around \quasar and which we also used to select a suitable narrow-band filter used for our CTIO observations. \citet{Kim_z1.123} estimated from the UVES spectrum $z=3.123$. A more recent study of the proximity effect in quasar spectra \citep{Proximity2008} quotes $z=3.120$ derived from C\,II line and $z=3.109$ measured from Si\,IV+O\,IV lines. \citet{Proximity2008_2} calculated the redshift $z=3.1257$ from the O\,I line. For our analysis we adopted $z=3.110$ which would place our quasar almost at the center of the narrow-band filter. In the case of a larger redshift the quasar would be shifted towards the red side of the filter and at the extreme value of $z=3.1257$ it would be located at the edge of the filter. We will return to this question later in Section~\ref{subsec: genconstraints}.

\begin{figure}[t!]
\includegraphics[width=0.5\textwidth]{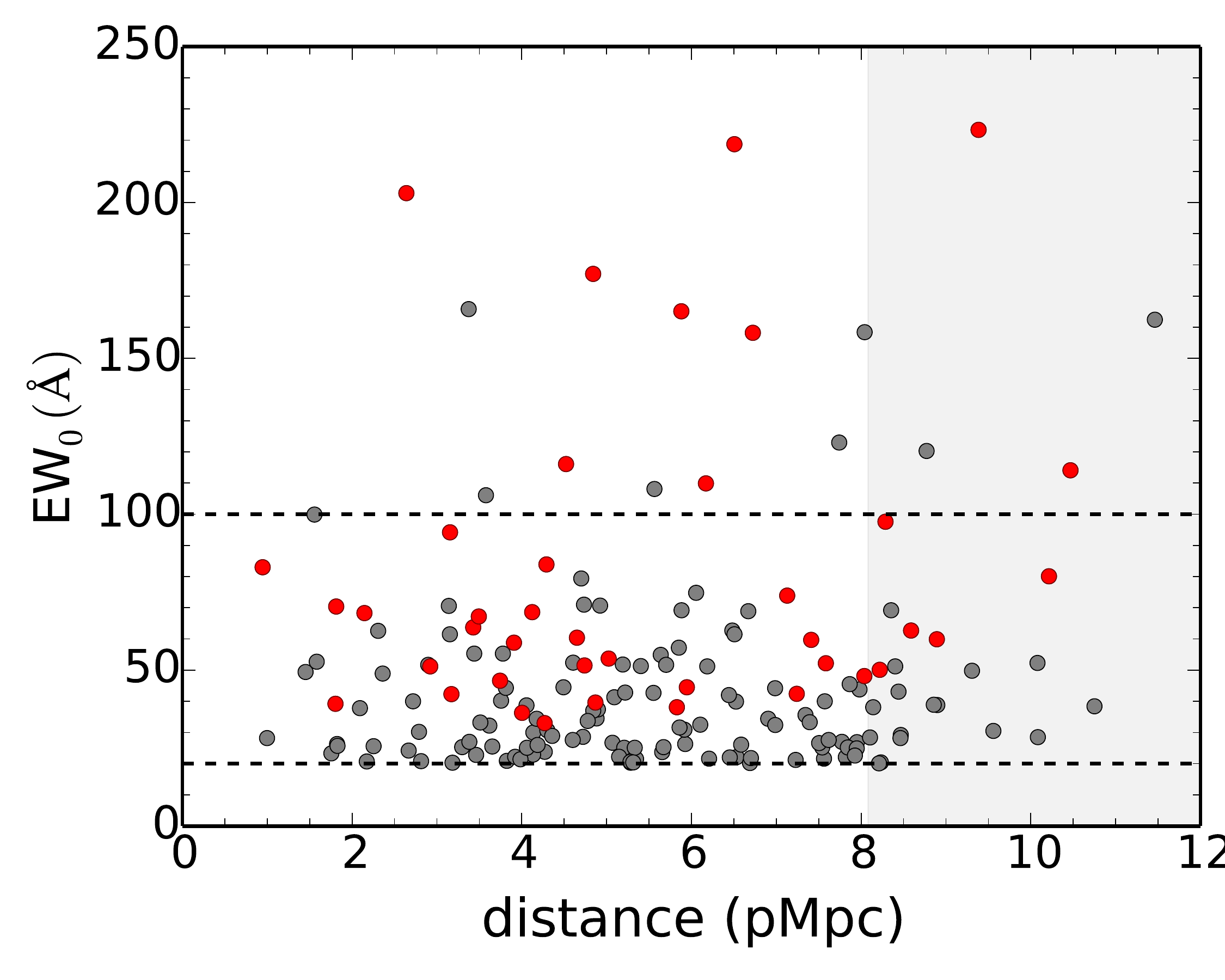}\caption{The distribution of projected distances from the quasar to LAE candidates as a function of rest frame equivalent width. The limit arrows for the red points are omitted in this figure.}
\label{distew}
\end{figure}

\subsection{Data Reduction}
\label{sec: datareduction}
The raw frames were reduced using the IRAF/MSCRED package which has been developed for cameras consisting of mosaic of CCD detectors. Each individual exposure was bias subtracted, flat-fielded and sky subtracted. Flat-fielding was done in two steps. First of all using the dome flats and then applying the night time flat, or superflat, produced by combining all the night time exposures together with the three sigma-clipping rejection algorithm to eliminate the objects. Calibration files were selected according to the date when the object was observed. Cosmic rays were removed from each individual exposure using LA Cosmic task~\citep{LACosmic}. The astrometric solution was found by matching the positions of stars in the field with the USNO-A2.0\footnote{http://www.usno.navy.mil/} catalog. Subsequently, all the individual reduced and co-registered exposures were projected onto the tangential plane using a single reference frame.

\begin{deluxetable}{rlrcr}
\centering
\tablecolumns{8}
\tablewidth{0pc}
\tabletypesize{\footnotesize}
\tablecaption{Characteristics of observations}
\tablehead{
           \colhead{Filter Name}                           & 
           \colhead{$\lambda_{cen}$\,\tablenotemark{a}}           & 
           \colhead{FWHM\,\tablenotemark{b}}           & 
           \colhead{Total integration\,\tablenotemark{c}} & 
           \colhead{m$_{3\sigma}$\,\tablenotemark{d}}       \\ 
           \colhead{}                                      & 
           \colhead{(\,$\angstrom$ )}                                      & 
           \colhead{(\,$\angstrom$ )}                                      & 
           \colhead{(hours)}                                      & 
           \colhead{}                                              
}   
\startdata
$u$                    &    3600   &   400     & $3$    & 25.539  \\
$g$                    &    4813   &   1537    & $1$    & 25.706  \\
$r$                    &    6287   &   1468    & $10$   & 26.054  \\
$[\mbox{OIII}]\,c6014$ &    4990   &   50      & $14$   & 25.751 
\enddata

\tablenotetext{a}{Central wavelength of each photometric filter.} 
\tablenotetext{b}{Full Width Half Maximum (FWHM) of each photometric filter.}
\tablenotetext{c}{Total integration time.} 
\tablenotetext{d}{The $3\sigma$ noise limit measured in 3 arcsec diameter aperture.} 
\label{filttable}
\end{deluxetable}

We used photometric standards observed during the same nights to do the flux calibration of the broad-band images. Each standard star exposure was reduced with the same pipeline and measured fluxes were converted to the AB photometric system using $ugr$ photometry for Landolt stars. In the case of the narrow-band filters, we used two spectrophotometric standards from \citet{Oke} to do calibrate the flux levels.

Flux-calibrated images were combined to produce a final stacked image in each filter. For the narrow-band image, which was used as the detection image, we produced stacks from all 29 exposures using average and median methods as well as two half-stacks with half of the exposures in each which were then used to crosscheck detected objects. The $3\sigma$ limits in magnitudes in our final stacks within a $3"$ diameter aperture in each of the photometric bands are shown in Table~\ref{filttable}.

\subsection{Catalog of \lya emitters}
\label{subsec:catalog}

We used {\it SExtractor}~\citep{Sextr} to detect objects in the narrow-band image. For the input we set the grid of detection thresholds varying from 1$\sigma$ to 2$\sigma$ of the local background and minimum detected areas varying from 4 to 10 pixels. Broadband fluxes of objects detected in the narrow-band image were measured in dual mode, i.e. with the centers of the objects and the aperture sizes defined from the narrow-band image. The error in relative astrometry between the different bands of $\sim0.24$\,pixel is small and allows us to use this approach. The photometric measurements were improved with a custom routine based on the IRAF/phot task. Noise properties were estimated using the approach described in \citet{Gawiser}. We have also carefully quantified the noise of each individual CCD in the mosaic. The variation between the best and the worst chip is $\sim20\%$. We have characterized the completeness level by adding point sources to the narrow-band image and running the detection algorithm which we used to search for \lya emitters. The value corresponding to $90\%$ completeness level across the whole image with the limiting magnitude is m$_{AB}=25.102$ and at this limit completeness for different detectors differed by about $10\,\%$ (i.e. between $85\%-96\%$).

The fluxes of the detected objects were measured in circular apertures of 3 arcsec diameter, except for a small number of very compact sources for which an aperture between 2 and 3 arcsec was used. 
We used these values to compute equivalent width (EW) measure which was used to select \lya candidates from our photometric catalogs. It was calculated according to the following formula, converted to the rest-frame assuming the lines were \lya at the redshift of the quasar:

$$
EW_0 = \left(\frac{F_{\mbox{\tiny NB}}}{F_{cont,\,\lambda_{\mbox{\tiny NB}}}} - 1\right) \frac{\Delta\lambda}{(1 + z_{qso})},
$$

\noindent where $F_{\mbox{\tiny NB}}$ is the narrow-band flux; $F_{cont,\,\lambda_{\mbox{\tiny NB}}}$ is the continuum flux at the central wavelength of the narrow-band filter; $\Delta\lambda$ is the FWHM of the narrow-band filter and $z_{qso}$ is the redshift of the quasar. As the flux limit in the $r$-band was significantly deeper than in the $g$-band and as this filter cannot be contaminated by \lya line emission at the redshift of the quasar, the $r$-band image was usually used to estimate the continuum flux of the objects. In the case with a secure detection in the $g$-band (above $3\sigma$), we calculated the slope of the continuum $f_{\lambda} \sim \lambda^{\beta}$ and interpolated the flux to the observed wavelength of the \lya line. The fluxes in the $g$-band were corrected for the line contribution following the approach described in \citet{Guaita2010}, where the spectrum was assumed to be flat with a narrow \lya line at the central wavelength of the narrow-band filter. If the flux in the $g$-band was below 3$\sigma$, the continuum was assumed to be flat in $f_{\nu}$, i.e. in frequency space. When the $r$-band flux was below $3\sigma$ we took the same approach as in \citet{C2012} using the 2$\sigma$ value as the upper limit on the continuum flux, or low limit on equivalent width. Objects with the rest-frame equivalent widths above 20\,$\angstrom$ were selected as candidates of \lya emission. As was discussed in previous works, this cut should produce less than 2\% [OII] interlopers at $z\approx0.3$ \citep{Gronwall}.

\subsection{High-EW \lya emitters}
\label{subsec: highew}

The final catalog of candidate \lya emitters (LAEs) contains 264 objects (Table 2 with flux measurements will be available in the published version of the paper). While the lower limit for the equivalent width was set at 20\,$\angstrom$ in the rest frame, the threshold to isolate putative fluorescent emission is not well defined. The limits for the strength of \lya emission have been studied using numerical simulations of stellar evolution. \citet{ew240_Charlot} have show that in the case of both constant and a burst of star-formation for different IMF parameter the upper limit for the \lya equivalent width is around $240\angstrom$. This limit was used by \citet{C2012} to identify dark galaxies in the proximity of a hyper-luminous quasar. \citet{Schaerer2003} has concluded that for metallicities between solar and $4\times10^{-4}$ the maximum equivalent width is in the range between 240-350$\angstrom$. According to \citet{ew240_Charlot}, there are strict requirements on the age of stellar populations (less than $10^8$ years) to show equivalent widths between $100$ and $200\angstrom$.  Values less than $100\angstrom$ are much more easily produced by stellar radiation.

From the observational point of view, star-forming galaxies typically show values of \lya equivalent width below $\sim100\angstrom$~\citep{Kornei2010}. This lower equivalent width cut was recently used by \citet{Trainor} to identify fluorescent boosting in the close vicinity of hyper-luminous quasars at the redshifts of $z\approx2.7$ and to infer lifetimes and opening angles from the distribution of the emitters around quasars.

In our sample of LAEs, we do not see any objects with equivalent width exceeding the more robust limit of $240\angstrom$. 
However, there are 16 objects with equivalent widths above $100\angstrom$, as can be seen in Fig.~\ref{magew}. The grey symbols show objects detected in the continuum and the red symbols correspond to the objects with the continuum levels below $3\,\sigma$ of the background. Following \citet{Trainor}, we have assumed that the sources with EW$_0\,> 100\angstrom$ are at least partially boosted by the quasar and can therefore in principle be used to obtain information on the quasar geometry and lifetime. We stress, however, that with this lower threshold our sample can be contaminated by the sources which have no contribution from \lya florescence. The narrow-band imaging does not allow us to distinguish between the contribution of other mechanisms, which are able to produce strong \lya emission, for example, presence of (attenuated) AGN or Population III stars \citep{ew240_Charlot,Schaerer2003}.

One of the possible (but not necessary) characteristics of boosted fluorescence is a relation between the distance from the quasar and the surface brightness and/or the EW. However, this holds only if the illuminated clouds are self-shielded to the incident ionizing radiation.  Other factors, which can also wash out this dependence, could be contributions from other sources of Ly a emission; uncertainty on the actual distance to the emitters in the space restricted by the width of the narrow-band filter.

In Fig.~\ref{distew} we show the distribution of the equivalent widths with respect to the projected physical distance from the quasar. In the selected sample of emitters we do not see expected decrease in the EW with the distance from the quasar. The objects with equivalent widths above $100\angstrom$ are found at all distances from the quasar except within an inner $2\,$pMpc region and the distribution of equivalent widths appears largely independent of projected distance from the quasar.

It should be pointed out that ``plausible candidates'' for fluorescent emission, with higher equivalent widths, have been discovered in the previous ``multi-slit'' spectroscopic study of this same field~\citep{C2007q}. Within a $3.4\,$pMpc region total of 13 \lya sources were found with equivalent widths, line profiles and the surface brightness suggested that half of them might be fluorescent. Most of these objects are detected in our catalog but have lower equivalent width limits than the nominal $100\angstrom$ cut for fluorescence. This is generally because these objects are faint (as the line emission were found spectroscopically) and do not have detectable continuum and thus their equivalent width measurements have relatively low limits. Some of them also fall off our narrow-band filter.

\begin{figure}[t]
\includegraphics[width=0.5\textwidth]{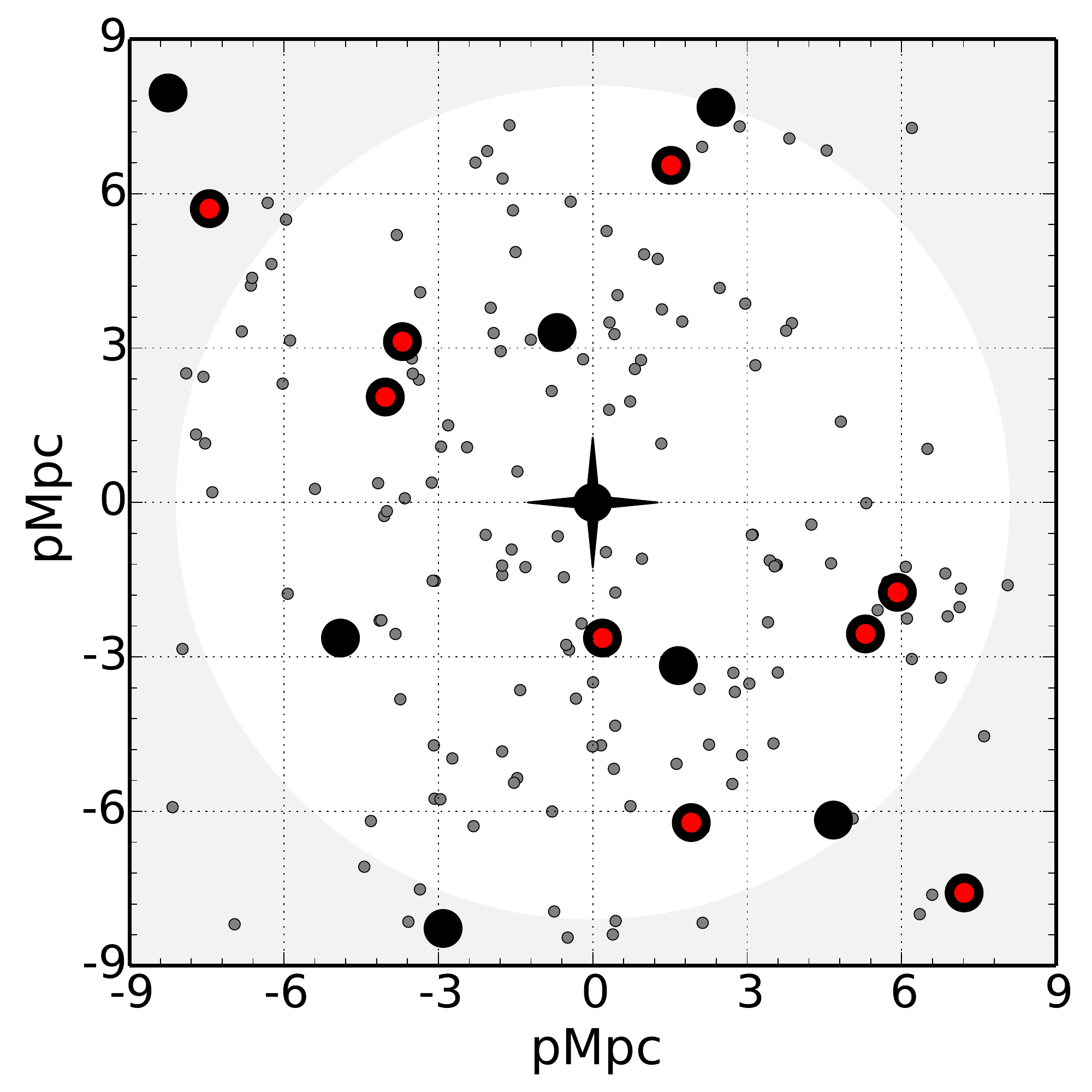}
\caption{ The distribution of emission line sources around the quasar in the plane of the sky. The highlighted symbols are objects with $EW_0 > 100\angstrom$ color-coded as in Fig.~\ref{magew}. The central symbol marks the position of the quasar. The white non-shaded area shows the circular region centered on the quasar used for some of the statistical tests.  If these high equivalent widths are indicative of fluorescent boosting by the quasar, then their existence at large projected distances sets significant constraints on quasar properties.}
\label{filed}
\end{figure}

\begin{figure}[t]
\includegraphics[width=0.5\textwidth]{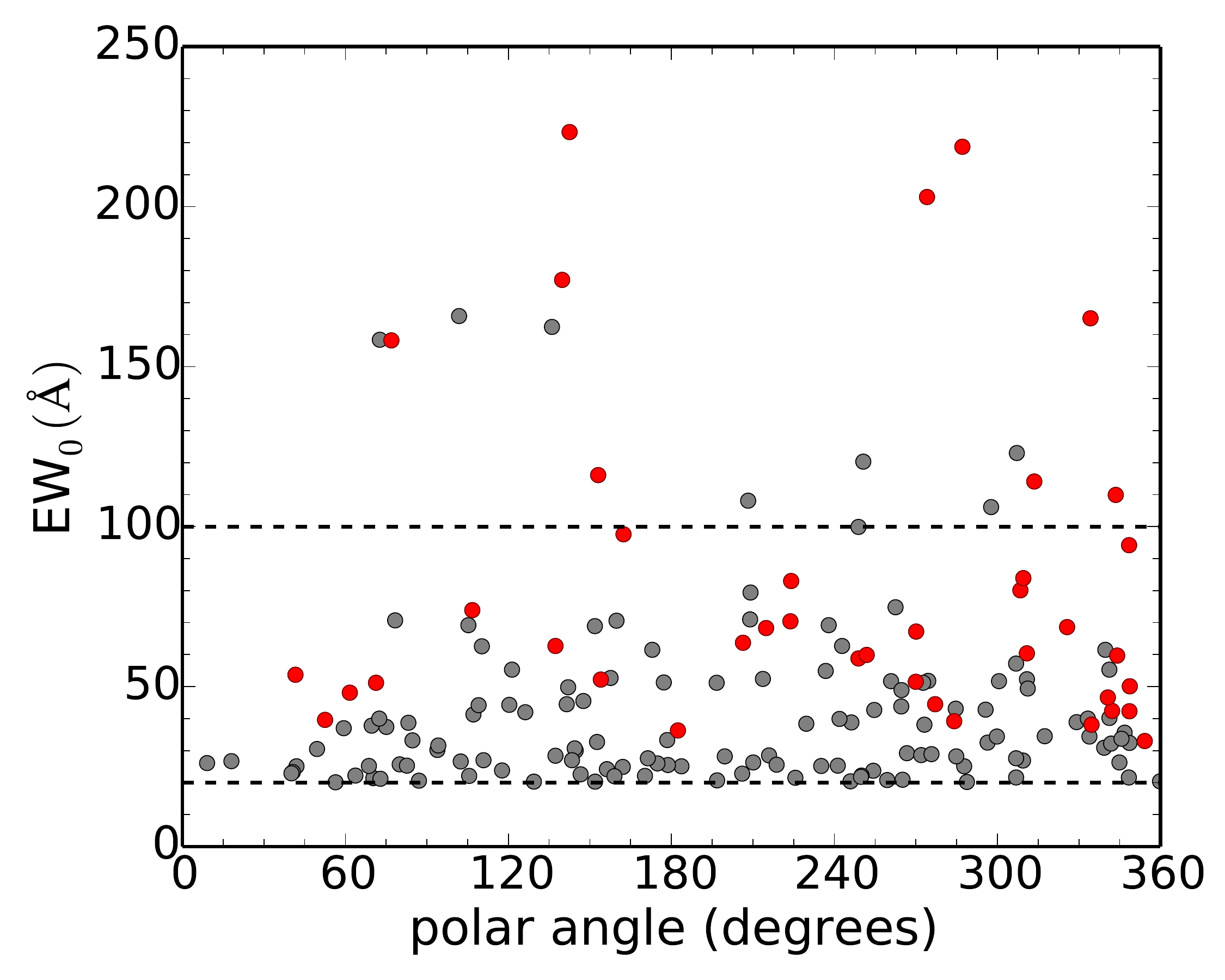}
\caption{The distribution of polar angles for the LAE candidates for different equivalent widths, color-coded as in Fig.~\ref{magew}.}
\label{angle-ew}
\end{figure}

\subsection{Spatial distribution of high-EW \lya emitters}

If the high-EW LAEs are being boosted by the quasar, then their spatial distribution may reflect the quasar lifetime and any anisotropy of the quasar radiation, as explored in Section 2. In Fig.~\ref{filed} we show the (x,y) distribution of objects in the field, highlighting the positions of the high equivalent width objects. The quasar is at the central position. 

Two things are apparent.  First, the high-EW LAE are found all the way out to the edges of the field with projected distances on the plane of the sky of upto about $60\,$cMpc.  If these most distant sources are being fluorescently boosted this fact immediately constrains the lifetime of the quasar, as described below. However, only a small fraction of the flux of these far away emitters can be explained by the quasar boosting unless the quasar was much brighter in the past. In particular, for an unresolved object with m$_{\mbox{\scriptsize AB}}\approx25^{m}$, measured in a 3 arcsec aperture, only about $13\%$ of the flux can be fluorescent if the luminosity of the quasar was constant, and the same in that direction as in ours \citep[see][]{C2007q}.

Second, the distribution on the sky shows an asymmetry with two distinct regions extending out on opposite sides of the quasar that contain very few high-EW sources. This can be seen in Fig.~\ref{angle-ew}, which presents the polar angles against the rest-frame equivalent width of all LAE candidates. The polar angle is defined between the horizontal axis and the vector connecting the quasar and the emitter anti-clockwise. It can be seen that high-EW LAEs occupy two regions of polar angle separated by two ``zones of avoidance'' with the $180^\circ$ gap.  In contrast, low-EW objects are distributed much more uniformly in polar angle.

We have explored statistically the significance of this azimuthal asymmetry. For this we were using the Kuiper test~\citep{Kuipertest,numres} which is analogous to the well-known Kolmogorov-Smirnov test but appropriate for a periodic variable such as angle. We compare the angular distribution of the high-EW sample with (a) a uniformly distributed sample within a circular region of radius $\sim10\,$pMpc and (b) with the distribution of the low-EW samples in both the same circular aperture and also the full square field. The statistical significance of the asymmetry is in all these tests below $2\,\sigma$, peaking around the $100\angstrom$ equivalent width cut.

\subsection{General constraints on the lifetime and opening angle}
\label{subsec: genconstraints}

We start by considering the constraints that would come from observing two fluorescent candidates at projected distances of $\sim15\,$pMpc on the opposite sides of the quasar.
By a simple geometrical argument, we can put lower limits on the opening angle and quasar lifetime.  

\begin{figure}[t!]
\includegraphics[width=0.49\textwidth]{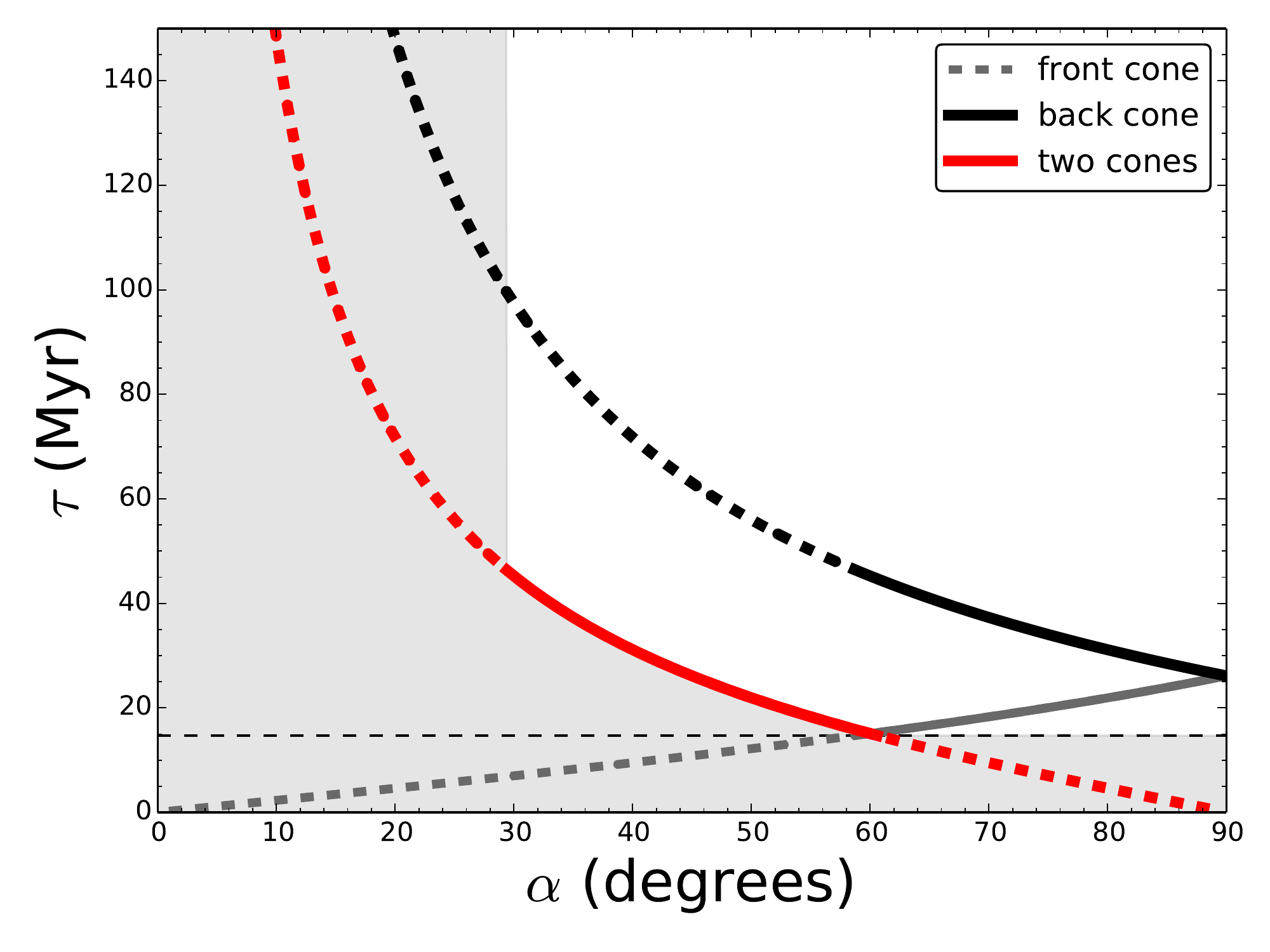}
 \caption{Constraints on quasar lifetime and opening angles that come from the two most distant LAEs if they are being fluorescently boosted by the quasar. Three different cases for the locations of the sources are considered - see text for details. The combined allowed parameter space is shown as non-shaded area.  A minimum quasar lifetime of $15\,$Myr is required, but this requires quite large opening angles. The minimum lifetime increases to about $40\,$Myr if the opening angle is reduced to $30\degr$.}
 \label{argument}
\end{figure}

We rewrite the equation for paraboloid for the time delay in terms of projected distance to the quasar line of sight and the angle $\theta$. We set $y=0$ for simplicity and consider the $(x,z)$ plane.

$$
c \tau = |r_{em}| + z_{em} = |x_{em}|\left( \frac{1+\cos \theta_{em}}{\sin \theta_{em}}\right)
$$

\noindent
where $r_{em}$ is the vector connecting the quasar and the position of a given Ly$\alpha$ emitter and $x_{em}, z_{em}$ its $x$ and $z$ components; $\theta_{em}$ is the angle between the line of sight and $r_{em}$, with $\theta_{em}=0\degr$ for an object behind the quasar and $\theta_{em}=180\degr$ for an object in front.  Three geometric configurations must be considered to put constraints on the lifetime $\tau$ and opening semi-angle $\alpha$, leading to three separate constraints in Fig.~\ref{argument}.

The first configuration is when the two objects are illuminated by different lobes of the bi-cone. Let the front object be ``1'' and the rear object be ``2''. In this case the constraint is set by the more distant object and the opening angle will be set by the larger of $(180-\theta_1)$ and $\theta_2$. The smallest opening angle requires a geometry when both objects have the minimum possible $(180-\theta_1)$ and $\theta_2$. This case is shown as the dashed and solid red lines in Fig.~\ref{argument}.

In the second configuration both objects are on the far side of the quasar and illuminated by the same cone. This is shown as a black line. The difference from the previous case is that the angles are now required to be twice as large because both sources must be included in the same illuminated cone.  This curve is therefore basically the same red curve but expanded by a factor of two along the (horizontal) opening angle axis. 

In the third configuration, both of the two objects are being illuminated by the same lobe of the quasar emission bi-cone but are located on the observer side of the quasar. This configuration allows the shortest lifetime of the quasar since the angle $\theta$ can in principle be arbitrarily small and is limited only by the finite width of the narrow-band filter effectively limiting the distance from the quasar. This locus is shown as the dashed gray line in Fig.~\ref{argument}. It ultimately sets the minimum required lifetime for the range of opening angles.

We did this analysis for the most optimal case when the quasar is located at the center of the narrow-band filter. As we discussed earlier, the redshift of the quasar is rather uncertain and the quasar might be shifted to the red side of the filter. That would involve a shift of the (allowed) solid lines along the dashed lines in the Fig.~\ref{argument}. If we consider the case of the two objects illuminated by different lobes of the radiation cone, the change of the position of the quasar towards the edge of the filter would allow smaller opening angles but larger lifetimes for the object in front of the quasar, and vice versa for the background emitter.

\begin{figure}[t!]
\includegraphics[width=0.5\textwidth]{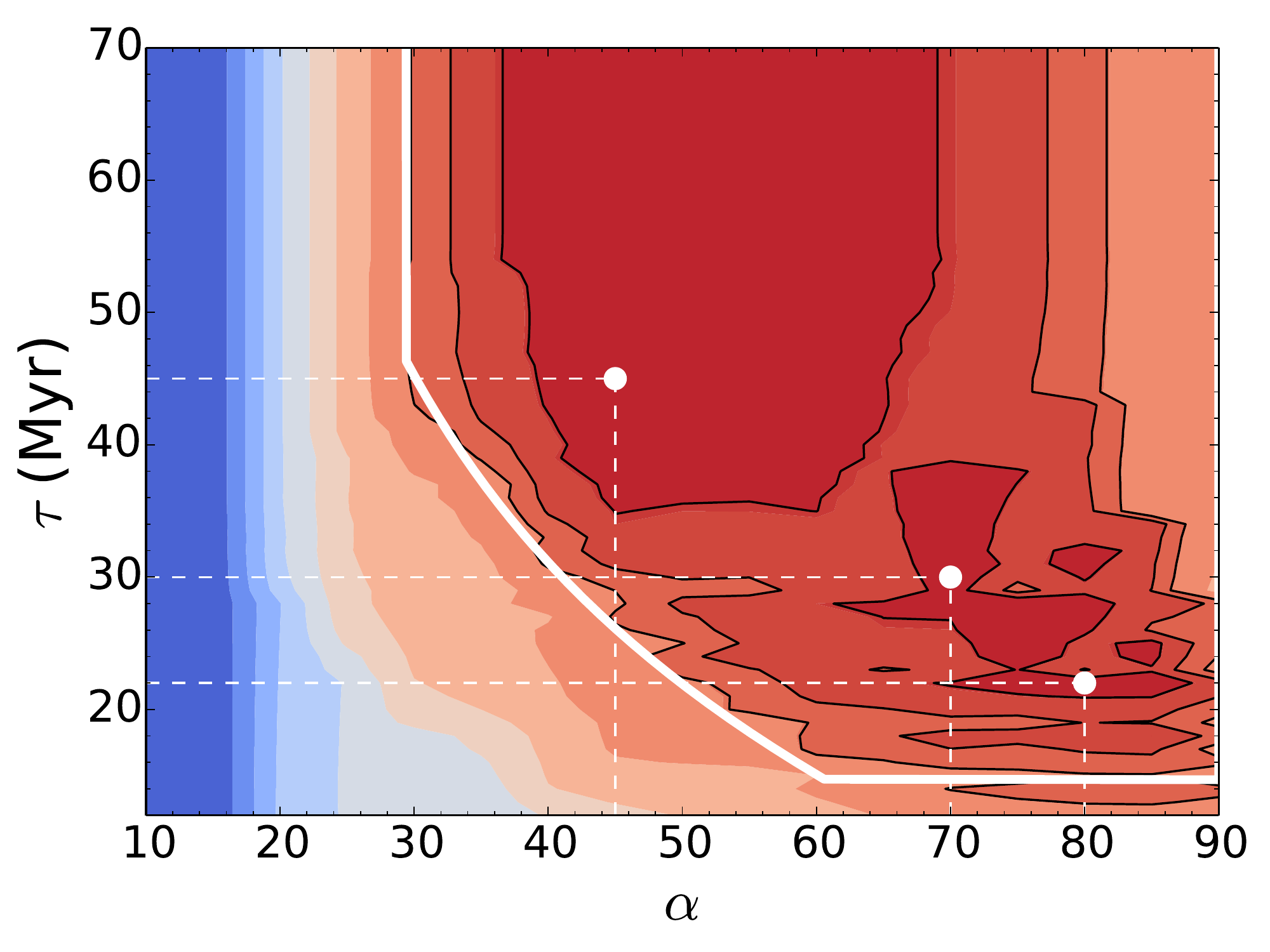}
 \caption{Parameter space for the best fit models estimated with the Maximum Likelihood approach. The contours correspond to e-fold drop in the likelihood function. The white dots are the show cases for density distribution as a function of redshift shown in Fig.~\ref{n-z}.}
 \label{mlhood}
\end{figure}

The white non-shaded area shows the resulting allowed region in parameter space if we do not have individual distance information (along the line of sight) for the fluorescent objects, as is the case here. Provided that our sample of high-EW objects are indeed being boosted by the quasar, we require $\tau > 15\,$Myr and $\alpha > 30\degr$, with more stringent constraints in combination. This also implies a range for the fluorescent boosting of about $60\,$cMpc.

\subsection{Best fit models: constraints on opening angle and lifetime from the full distribution}
\label{subsec: maxlikelihood}

Using our toy model we then built an extensive library of models for a range of input parameters to compare it with the (x,y) distribution of the data. For simplicity, the position angle in the plane of the sky was fixed to align the ``zone of avoidance'' with the minima in the azimuthal distribution of high-EW objects in Fig.~\ref{angle-ew}. Opening angles in the range $0^\circ~<~\alpha~<~90^\circ$ and viewing angle $0^\circ~<~\psi~<~\alpha$ (i.e. from the case when the observer is looking down the axis of the cone to the case when the observer is looking along the surface of the illumination cone) were explored.

We used a Maximum Likelihood method to select acceptable models, taking the projected density field of each model as a probability of seeing a fluorescent object at this position. In each case, the density field was normalized so that the total expected number of sources matched the size of the observational sample. To avoid the Likelihood being driven by one or more sources lying in regions outside of the region illuminated by the quasar, we assigned these blank regions a uniform probability density corresponding to a 1\% chance of finding an object there.   Such situations would be strongly discouraged, but not excluded at a first place.

\begin{figure}[t!]
\includegraphics[width=0.5\textwidth]{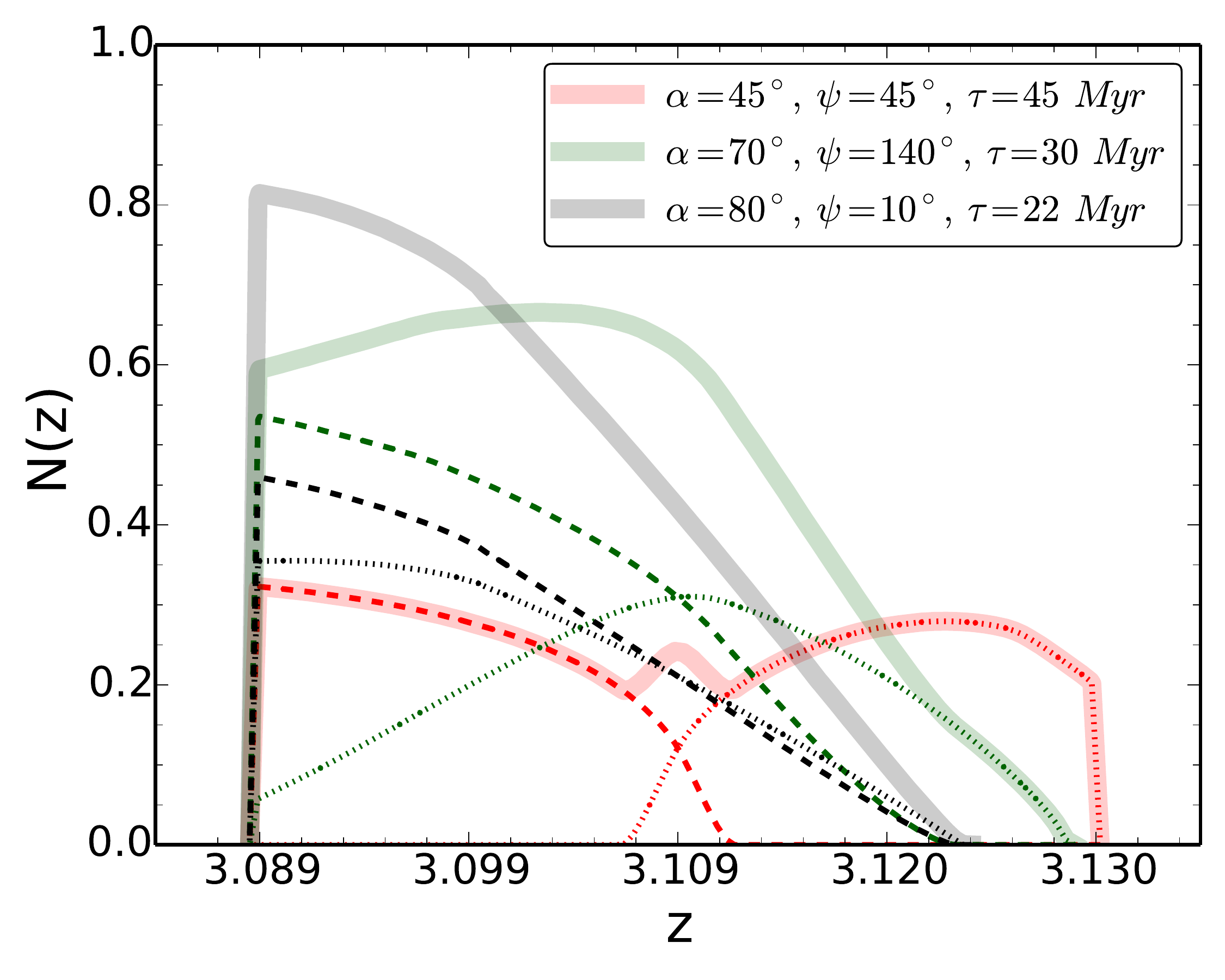}
 \caption{Number density of objects illuminated by quasar radiation as a function of redshift. Each color corresponds to one of the three best fit models in Fig.~\ref{mlhood}. The dotted and dashed lines show densities for each of the lobes of the two-cone model and the solid line is the summed density value. Different models can be tested with high precision redshift observations.}
 \label{n-z}
\end{figure}

\begin{figure}[t!]
\includegraphics[width=0.5\textwidth]{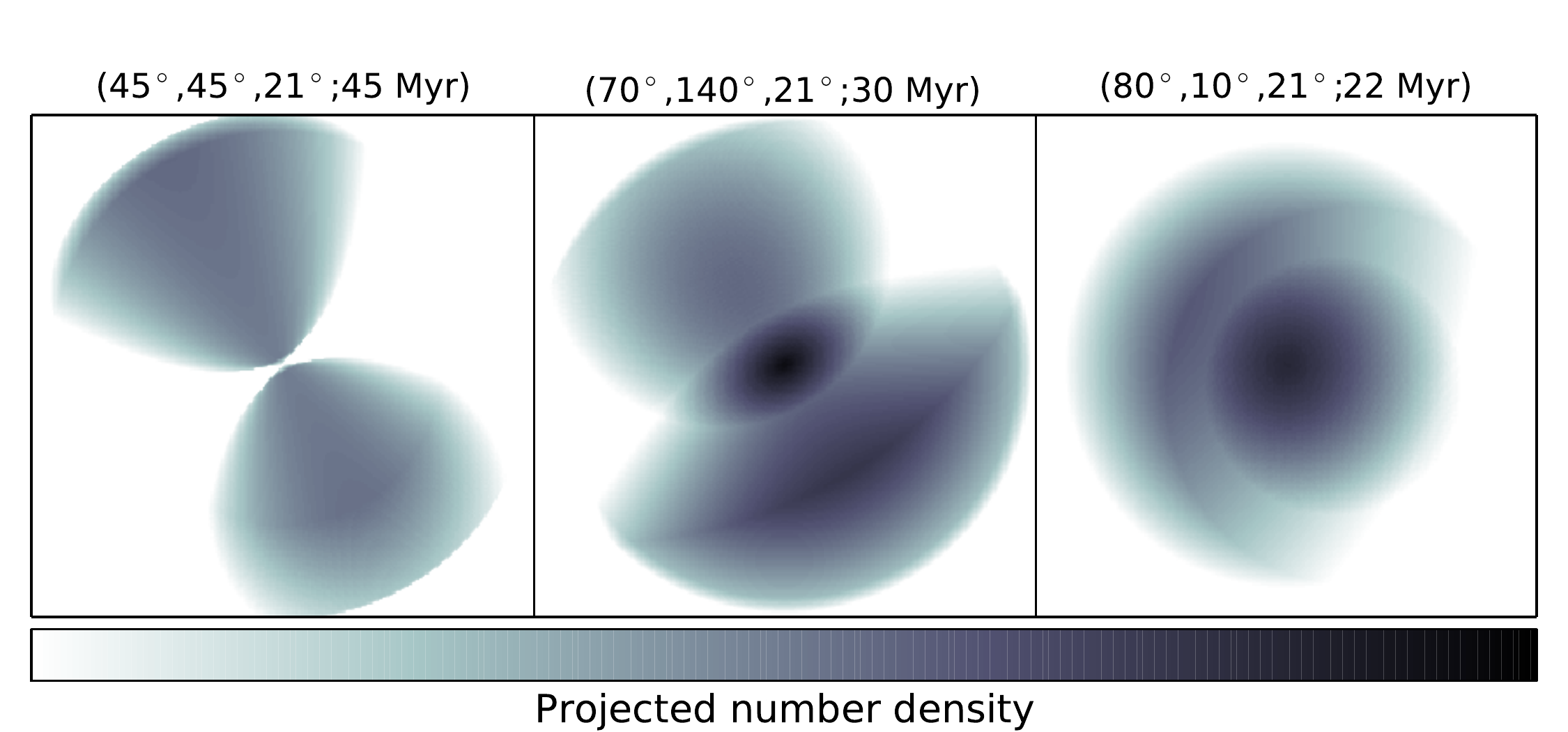}
 \caption{Projected density of objects illuminated by the quasar for different best fit models from Figures~\ref{mlhood}. The caption on the top of each subplot indicates the parameters of each model ($\alpha , \psi , \phi$; $\tau$).}.
 \label{n-z-3D}
\end{figure}

The results are presented in Fig.~\ref{mlhood}. The inside contours correspond to (1,2,3)-$\sigma$ levels. As we can see, the shape of the likelihood function is very similar to the conclusions drawn from our basic geometry arguments which were shown in Fig.~\ref{argument} . The best models are those for larger lifetimes in case of relatively small opening angles and allowing shorter lifetimes in the case of larger opening angles. This is consistent with the conclusion reached by~\citet{C2007q} that the presence of the best fluorescence candidate behind the quasar implies $\tau\ge60\,Myr$.

These results are different from the \citet{Trainor} work who found for the sample of eight quasars at the redshifts $z\sim2.7$ the lifetimes in the range $0 < \tau < 20\,$Myr.  Their analysis was based on asymmetries in the redshift space distribution of emitters within much smaller projected distances (up to $2\,$pMpc) than those investigated (without redshift information) in the present work.   

Our models can be used as predictions for the redshift distribution of the objects illuminated by the quasar. The redshift distribution of LAEs for different opening angles, orientation and lifetimes can be used to distinguish between different models. This is shown in Fig.~\ref{n-z}. Three cases cover the range of the best fit models from the likelihood analysis. Dotted and dashed lines correspond to different lobes of the two-cone emission. The solid lines show the integrated density of the objects. For the smallest opening angle model the time delay is longer therefore the redshift distribution is relatively uniform. For the larger opening angles smaller lifetimes are allowed. In this case there is a greater asymmetry in the redshift distribution of objects. The 3D visualizations for the best fit models from Fig.~\ref{n-z} are shown in Fig.~\ref{n-z-3D}.

\subsection{Fluorescent or not?}

The constraints on quasar lifetime clearly come from the assumption that the \lya emission from the sources in question, selected to have EW$_0>100\angstrom$, is fluorescently re-radiated ionizing radiation from the quasar.  The case for this is by no means established, and there are a number of features of the data that argue for some caution in this association.  Specifically, we do not see the radial dependence on \lya luminosity or equivalent width that would be expected if the gas is self-shielded.  Also, the most distant sources, which drive the life-time constraint, are sufficiently distant that the quasar would have had to have been brighter in the past, or to be brighter in that direction than in the direction towards us, if the observed \lya emission was to be fluorescent.

If these relatively low EW$_0$ sources are not fluorescent and are instead produced by other mechanisms, then this would invalidate the application of the model to these sources to derive lifetime constraints on the quasar.  This would however, cast some question on the use of sources selected in the same way for the same purpose in \citet{Trainor}.

\section{Conclusions}

We have developed a toy model to explore the expected projected distribution of fluorescently illuminated sources in the vicinity of a bright quasar and to see how this depends on the quasar emission opening angle and lifetime, on the orientation of the system with respect to the line of sight and on the radial extent of the boosted region, and also on observational factors such as the width of a narrow-band filter used to identify the sources.    

We have then compared these model distributions to the observed distribution of high equivalent width \lya emitters (selected to show equivalent widths above $100\angstrom$, the limit used in the previous study by \citet{Trainor}) within a large 36x36 arcmin$^2$ field centered on the bright quasar \quasar at $z=3.110$. Our main conclusions are:

\begin{enumerate}

 \item{Information of quasar opening angles and lifetimes, estimated from flourescent \lya emission, comes from the maximum projected distance of fluorescently emitting objects from the quasar, from possible azimuthal asymmetries in their distribution and especially the possible existence of narrow ``zones of avoidance'', and from apparent asymmetries between the two lobes. }

 \item{The emission line objects in the \quasar field that show equivalent widths above $100\,\angstrom$ extend across the whole field with projected separations up to $15\,$pMpc. There is some evidence that these are distributed in two lobes that are separated by a narrow ``zone of avoidance'' of the type expected for anisotropic bi-conic quasar emission.}

\item{On the assumption that these high equivalent width objects are fluorescent emitters with \lya emission boosted by the quasar emission, as assumed also by Trainor \& Steidel (2013) in their similar study, we have applied our model to derive constraints on the opening angle and lifetime of the quasar emission.  The existence of two emission line sources at high projected distances of $15\,$pMpc requires a lifetime of at least $15\,$Myr for an opening angle of $60\degr$ or more, increasing to more than $40\,$Myr if the opening angle is reduced to a minimum $30\degr$. The overall distribution of these sources across the field gives best fit lifetimes in the range $20 < \tau < 50\,$Myr for opening angles in the range $90\degr < \alpha < 40\degr$.}

\item{However, the presence of these emitters at such large distances would, given the present brightness of the quasar, require that the quasar was significantly brighter in the past or in the direction of the sources in question.  Furthermore, the population of sources do not show some of the expected signatures at the current survey limits, including a radial fall off in \lya luminosity or equivalent width expected for self-shielded gas.  These may both argue against a fluorescent origin for most if not all of the \lya emission in these sources.
}

\item{If the sources with equivalent width above $100\angstrom$ are indeed being fluorescently boosted by quasar radiation, then these observations place significant constraints on quasar properties. If not, then it suggests that a much higher EW limit, then the one used by \citet{Trainor}, e.g. the $240\angstrom$ limit used by \citet{C2012}, must be used to isolate fluorescence and that results with this lower threshold should be treated with some caution.
}

\end{enumerate}

To summarise, the distribution of \lya fluorescent emitters, boosted by quasar radiation, can in principle be used to estimate quasar lifetimes together with the geometry of its radiation. However, this method encounters some challenges and at this point cannot be considered as robust. 
On the modelling side, the toy model could be improved to include, for example, the expected surface brightness of boosted \lya emitters based on the quasar luminosity, the decline of the boosting effect with distance, and the possible delays in the \lya due to recombination time, etc.

From the observational point of view, the most important question is how can we confidently select fluorescent \lya emitters.
As discussed in the paper, the lower equivalent width cut of $100\,\angstrom$ may mean that the sample of \lya emitters is contaminated by non-fluorescent objects. Many of the fainter objects in this paper only have lower limits to their EW, and could well in fact have EW $> 240\,\angstrom$.  However, establishing this requires very deep continuum measurements. Furthermore, if the redshift is not known, the use of narrow-band filters also introduces systematic uncertainties in the line measurements, biassing the equivalent widths lower, due to the filter profile. Another possible contamination of very high equivalent width objects is from AGN.  These could be recognised using very deep X-ray data.

Another potential problem with this technique, for long quasar lifetimes $> 10^7$ years, is illustrated by the current analysis. Even the brightest quasars may have zones of influence, i.e. the distance within which they can significantly boost fluorescent emission, that are smaller than the distance light travels in the available time.  In other words, the distribution of detectable signatures of fluorescence may be limited by the available light rather than the available time. \vspace{0.5cm}

\section*{Acknowledgements}
We thank the referee for helpful comments on the paper which improved the content and presentation of this paper. This work has been supported by the Swiss National Science Foundation. It is based on observations at Cerro Tololo Inter-American Observatory, National Optical Astronomy Observatory (NOAO Prop. ID 2011B-0407, PI: Gabor Worseck), which is operated by the Association of Universities for Research in Astronomy (AURA) under a cooperative agreement with the National Science Foundation. JXP acknowledges support from the National Science Foundation (NSF) grants AST-1010004 and AST-1412981.

\bibliographystyle{apj}
\bibliography{apj-jour,refs}

\end{document}